# Epigenome-wide association study and integrative analysis with the transcriptome based on GWAS summary statistics

Hon-Cheong So[1,2*]

[1]School of Biomedical Sciences, The Chinese University of Hong Kong, Shatin, Hong Kong

[2]KIZ-CUHK Joint Laboratory of Bioresources and Molecular Research of Common Diseases, Kunming Institute of Zoology and The Chinese University of Hong Kong

*Email: hcso@cuhk.edu.hk

**Abstract**

The past decade has seen a rapid growth in omics technologies. Genome-wide association studies (GWAS) have uncovered susceptibility variants for a variety of complex traits. However, the functional significance of most discovered variants are still not fully understood. On the other hand, there is increasing interest in exploring the role of epigenetic variations such as DNA methylation in disease pathogenesis. In this work, we present a general framework for epigenome-wide association study and integrative analysis with the transcriptome based on GWAS summary statistics and data from methylation and expression quantitative trait loci (QTL) studies. The framework is based on Mendelian randomization, which is much less vulnerable to confounding and reverse causation compared to conventional studies. The framework was applied to five complex diseases. We first identified loci that are differentially methylated due to genetic variations, and then developed several approaches for joint testing with the GWAS-imputed transcriptome. We discovered a number of novel candidate genes that are not implicated in the original GWAS studies. We also observed strong evidence (lowest $p$ = 2.01e-184) for differential expression among the top genes mapped to methylation loci. The framework proposed here opens a new way of analyzing GWAS summary data and will be useful for gaining deeper insight into disease mechanisms.

## Introduction

The past decade has seen a rapid growth in omics technologies. Genome-wide association studies (GWAS) have been successful in unraveling susceptibility variants for a variety of complex traits[1]. However, the functional significance of most discovered variants are still not fully understood. On the other hand, high-throughput technologies have enabled the study of gene expression and epigenetic changes on a genome-wide scale. Integrative analyses of these various types of data will deepen our understanding of disease pathophysiologies and ultimately lead to more effective treatments.

DNA methylation is a key epigenetic mechanism which is implicated in the regulation of gene expression[2]. Recent studies also suggested that DNA methylation may be involved in other functions such as promotor usage and alternative splicing[2,3]. With the recent advent of new high-throughput arrays, DNA methylation can now be studied genome-wide in a hypothesis-free manner. Such epigenome-wide or methylome-wide association studies (EWAS/MWAS) has been applied to study numerous diseases and many methylation sites have been found to be associated with disease status[4]. Nevertheless, the interpretation of EWAS findings is often not straightforward. Reverse causation is possible, that is, methylation changes can be the direct or indirect consequences of a disease instead of being its cause[5]. In addition, associations between methylation and disease can be due to confounding factors, such as smoking and other unknown variables that could not be completely adjusted for[4].

In view of these limitations, in this study we present a Mendelian randomization (MR) based analytic framework to study the relationships between methylation and disease. A genetic proxy is used to represent methylation and its association with disease is then investigated. MR is an established method to study causal relationships between exposures and outcomes[6]. MR has been previously proposed as a way to study the relationships of epigenetic processes to disease[5], but to our knowledge no studies have applied this approach in a methylome-wide level. In the paper the authors suggested a more complex two-step design, in which one first uses a genetic proxy for an environmental exposure (e.g. smoking) to assess the causal relationship between exposure and disease, and then uses another genetic proxy for methylation to elucidate the relationship between methylation and disease[5]. In this work we focus on the one-step approach. In essence, we studied the genetically controlled component of methylation and assessed the association of methylation levels with complex traits or diseases. We note that an association can be due to causality (i.e. methylation alterations causing the disease), pleiotropy [i.e. the proxy genetic variant(s) independently affect methylation and disease risks], or a combination of both. As will be discussed in a later section, we do not focus on distinguishing these possibilities in this work.

The MR approach for methylation studies is less vulnerable to confounding and reverse causation than conventional methods[6,7]. And importantly, this approach only requires summary statistics from GWAS and methylation quantitative trait loci (meQTL) studies. With the increased availability of GWAS summary results (e.g. in LD Hub[8]) and rising sample sizes from meta-analyses, this approach leverages the power of GWAS to unravel methylation changes linked to complex diseases. To our knowledge, this is the first study to investigate methylome-wide associations based on GWAS summary data.

We developed a framework to integrate the methylation results with the GWAS-imputed transcriptome, and proposed several measures to assess the significance of a gene based on a joint analysis of GWAS-derived methylome and transcriptome associations. We applied the approach to five complex diseases and found a remarkable tendency (lowest $p$ = 2.01e-184) towards differential expression within the top methylation loci. We also revealed a number of novel genes for the diseases under study.

A few previous studies have interrogated alternative omics profiles from GWAS data. The majority focused on the transcriptome, for example Gamazon et al.[9] proposed imputation of expression levels based on training in a reference transcriptome dataset. The method was extended to the use of only summary statistics for imputation, including TWAS[10] and MetaXcan[11]. These two methods employed mixed effects models and elastic net regression respectively by default to train prediction models. However, in this study we relied on top summary statistics from meQTL studies (instead of having the original genotype data) and MR is a computationally simple and flexible approach that does not require training with raw data. In a related work, Zhu et al. proposed SMR[12], a MR-based approach which integrated GWAS summary data with expression QTL results. We studied the methylome in this work instead and compared to SMR, we did not restrict the analyses to only the top QTL in each test. We also extended our analysis to integrate methylome and transcriptome data.

**Methods**

**The Mendelian Randomization approach**

Mendelian randomization (MR) involves the use of genetic variants as instruments to interrogate the relationship between an exposure and an outcome[6]. Genetic variation serves as a natural way of "randomization" of individuals such that confounding is much less likely. Details of different MR methods could be found elsewhere[7]. In brief, assume that we have $j$ genetic variants ($G_1$, $G_2$…$G_j$) acting as instrument variables, $X$ is the exposure (risk factor), $Y$ is the outcome, $\beta_{Xj}$ is effect size of $G_j$ to $X$ and $\beta_{Yj}$ is the effect of $G_j$ to $Y$. The causal effect of $X$ on $Y$ can be estimated by $\hat{\beta}_{IVj} = \hat{\beta}_{Yj} / \hat{\beta}_{Xj}$ under certain assumptions. Note that $X$ refers to methylation levels and $Y$ refers to the studied disease or trait in our case. The assumptions include: (1) The genetic variant is independent of confounders; (2) The genetic variants is associated with the exposure; and (3) The genetic variant have no effect on the outcome when the exposure remains fixed, i.e. no pleiotropic effects. It is often difficult to fully exclude pleiotropy in actual scenarios. In the presence of pleiotropy, the effect size becomes $\beta_j = \beta + \alpha_j / \beta_{X_j}$, where $\alpha_j$ is the effect size of the pleotropic effects[13] and $\beta$ is the causal effect. Recently new methods such as Egger regression[13] have been developed to produce consistent casual effect estimates when pleiotropy is present. However, in this study the number of instrumental variants is small (mostly less than three) and hence we did not focus on separating causal effects from pleiotropy.

By the delta method, the variance of the effect estimate of *X* on *Y* based on an instrumental variable $IV_j$ is

$$\text{var}(\hat{\beta}_{IVj}) = \frac{\sigma_{Yj}^2}{\hat{\beta}_{Xj}^2} + \frac{\hat{\beta}_{Yj}^2 \sigma_{Xj}^2}{\hat{\beta}_{Xj}^4} \quad \text{-----------------(1)}$$

up to second-order terms[14]. In the inverse-variance weighted method, only the first-order term is retained[15]. We kept the second-order term in this study and combined the effects of different genetic variants by a fixed-effects approach. The aggregate effect size estimate of *X* on *Y* (*i.e.* methylation on disease state or trait values) can be expressed by:

$$\hat{\beta}_{agg} = \frac{\sum \hat{\beta}_{IVj} \text{var}(\hat{\beta}_{IVj})^{-1}}{\sum \text{var}(\hat{\beta}_{IVj})^{-1}}$$

and the corresponding variance is

$$\text{var}(\hat{\beta}_{agg}) = \frac{1}{\sum \text{var}(\hat{\beta}_{IVj})^{-1}}$$

We used the implementation in the R package MR-base[16]. The above algorithm was repeated for each methylation probe available in the meQTL dataset and repeated for each time-point (birth, childhood, adolescence and middle age).

**GWAS and methylation QTL data**

We performed methylome-wide association studies based on MR for five complex diseases, namely schizophrenia (SCZ)[17], Alzheimer's disease (AD)[18], coronary artery disease (CAD)[19], type 2 diabetes mellitus (DM)[20] and inflammatory bowel disease (IBD)[21]. GWAS summary statistics were derived from meta-analyses of large sample sizes and study details were given in the attached references. For methylation data, we extracted the results from a recent methylation QTL study known as ARIES[22,23]. The study investigated the influence of genetic variations on methylation from cord blood or peripheral blood samples across the life course by the Illumina Infinium HumanMethylation450 array[23]. They identified methylation QTLs (meQTLs) at five life stages, namely birth, childhood (mean age 7.5 years), adolescence (mean age 17.1 years), pregnancy (mean age 29.2 years) and middle age (mean age 47.5 years). The sample size was relatively large ($N$ ranging from 742 to 837). The results are publicly available and the independent top meQTL results (available at http://www.mqtldb.org/ and in the MR-base package[16]) were used in our study.

For SCZ and IBD, the median ages at onset are well below middle age[24,25] and hence we only studied methylation at three time-points (birth, childhood and adolescence) while four time-points were considered for the other traits.

**Joint analysis with GWAS-imputed transcriptome**

As methylation can affect gene expression and the transcriptome is an integral part of disease pathogenesis, in the next step we imputed transcriptome based on GWAS summary results. We employed MetaXcan[11] to impute expression changes in peripheral blood, in order to match with the tissue studied in the methylation QTL study. The DGN dataset[26] was used as the reference transcriptome dataset. MetaXcan pre-trains a prediction model (elastic net by default) on the reference set and applies it to the summary GWAS results.

To jointly analyze methylome and transcriptome data, we adopted a gene-based approach. Methylation probes were annotated based on the documentation provided by the Illumina HumanMap450 manifest files. For multiple probes that mapped to the same gene, we employed the Simes method to combine the p-values. The Simes method tests for at least one non-null effect among all hypotheses and controls the type I error rate under positive regression dependency[27]. After obtaining a gene-based *p*-value for methylation, we further combined it with the expression *p*-value from MetaXcan.

We considered two types of hypotheses in this case. In the first scenario, the researcher may be interested in finding genes that are *both* differentially methylated and expressed. We proposed a simple hypothesis testing approach by taking the maximum p-value i.e. *max*($p_{\text{methylation}}$, $p_{\text{expression}}$), which is proven to be valid under positive dependence[28,29]. In the second scenario, one may be interested to search for genes that are associated with disease in *at least one* kind of omics measure (expression or methylation). To put it in another way, the null hypothesis in the second scenario is the *global null*. The Simes method was used in this case to combine p-values.

For the second scenario, we also developed an empirical Bayes approach, in line with the local false discovery rate (fdr) framework[30] but adapted to the two-dimensional case. Here the null hypothesis ($H_0$) is that both effects are null. Let $z_1$ and $z_2$ be the observed z-statistics corresponding to differential methylation and expression, we define the "co-fdr" as:

$$co\text{-}fdr = \Pr(H_0 | Z_1 = z_1, Z_2 = z_2) = \frac{f(z_1, z_2 | H_0)\Pr(H_0)}{f(z_1, z_2)}$$

Here $f(z_1, z_2 | H_0)$ is the probability density function (pdf) of $z_1$ and $z_2$ under the null, which was assumed to follow a bivariate normal distribution with unknown means and covariance. We estimated the unknown parameters by fitting a truncated bivariate normal distribution to the observed z-statistics, and obtained the maximum likelihood (MLE) estimates. By doing so we assumed that the central part of the observed distribution of z-statistics are mostly null, and a similar approach has been used by Efron to estimate the empirical null for one-dimensional z-statistics[31]. Following Efron, we set the cutoff z-value at 2, i.e. we used z-values between -2 and 2 to fit the truncated normal distribution. We set $\Pr(H_0)$ (prior probability of null association) to be one to produce a conservative estimate of co-fdr. $f(z_1, z_2)$ was estimated by a multivariate kernel density function using the R package

"ks"[32].

**Differential expression among the top methylation loci**

Within those methylation loci that passed the Bonferroni correction (for the total number of probes tested at each time-point), we tested whether the genes corresponding to these top methylation loci were also *differentially expressed* as a whole. We considered the genes that were mapped to at least one methylation loci *and* had expression test results available. We collected the expression p-values of the top differentially methylated genes and combined them with the Fisher's method and the Simes method.

**Methylation-expression associations among the top methylation loci**

Among the top methylation loci, we also tested for Pearson and Spearman correlations of the effect sizes of differential methylation and expression. As previous studies showed that the methylation-expression link is associated with the location of methylation[2], methylation probes were grouped by both gene names and their locations (gene body, TSS1500 [1500 bp upstream of transcription start site], TSS200, first exon, 5' or 3' UTR) as annotated by the Illumina documentation file. If a gene was represented by multiple methylation probes or at different time-points, the mean effect size was used.

**Results**

**Top methylation loci identified**

We observed a high correlation among the test results that overlapped across different life stages, consistent with the report by Gaunt et al. in the original meQTL study. We mapped all genome-wide significant SNPs in GWAS summary data (p < 5e-8) to genes using the R package Biomart, and compared them with the genes corresponding to our top methylation loci. Table 1 shows the methylation loci which passed the Bonferroni correction (for the total number of loci tested at each time-point) and *not* implicated in the original disease GWAS. These results represent novel candidate genes and might warrant further follow-up studies. Note that the Bonferroni correction is conservative as there are correlations among the methylation probes.

**Joint testing with imputed transcriptome**

The top conjunction test results of methylation and expression using maximum *p* are shown in Table 2. Only the Bonferroni significant results are shown. It is worth noting that the maximum p method is simple and statistically valid but it does not consider the actual covariance structure so some power is lost in this regard. There are fewer results passing Bonferroni correction which could be due to inadequate power to detect such associations. However, genes showing evidence of both differential methylation and expression might represent more biologically plausible candidates for further functional studies.

Table 3 shows the top 10 genes with the lowest co-fdr for each trait in our analysis. The co-fdr can be directly interpreted as the probability that the gene shows no differential expression or methylation from the GWAS-based analysis. Both the co-fdr and Simes method test the global null hypothesis; hence low co-fdr tends to be associated with low Simes p-values and vice versa.

**Differential expression and methylation-expression associations among the top methylation loci**

We found that genes mapped to the top methylation loci (passing Bonferroni significance) also tended to be differentially expressed. The associations were remarkably strong, with the lowest p-value achieving 2.01e-184 (for IBD) (Table 4). The associations were also observed for all five traits under study. However we note that not all genes demonstrated such an association. One possible explanation for the enrichment of differential expression is that differential methylation results in transcriptomic changes, which in turns affect disease risk. This is biologically plausible given the evidence for methylation regulating expression[33]. However, we caution that our model cannot prove a causal relationship between methylation and expression. The reverse is also possible (i.e. expression changes leading to differences in methylation levels)[34]. It is also possible that the same genetic variant(s) affect methylation, expression and disease risk separately and we will still observe both differential methylation and expression with respect to the disease. Hence one explanation for our observations is that disease susceptibility variants tend to influence gene expression *and* methylation together, more than expected by chance. Given the complexity of biological systems, it is possible the truth may involve a mixture of the above mechanisms, and may vary for individual genes and disease phenotypes.

We also examined correlations of effect sizes with regards to methylation and expression among the top differentially methylated loci. For schizophrenia, we observed a significant negative correlation (Pearson correlation -0.571, *p* = 1.79e-4; Spearman's rho -0.522, p = 7.74e-4) (Table 5). We then performed a stratified analysis considering only methylation probes at the promotor region, which include TSS200 (200 base pairs away from the 5' end of transcription start site), TSS1500 (regions further away than TSS200 by up to 1300 base pairs), 5'UTR (5'untranslated region) and the first exon. Methylation in these regions was previously reported to be associated with repression of gene expression[35]. For schizophrenia, a negative correlation of effect sizes of methylation and expression was observed (Pearson correlation -0.61, *p* = 5.46e-3; Spearman's rho -0.63, p = 4.71e-3) and a binomial test showed enrichment for associations in opposite signs (opposite signs in 15/19, one-tailed p = 0.010). For methylation probes at other regions, we observed a negative Pearson correlation of effect sizes but the association was weaker (Pearson correlation -0.54, p = 0.017); and no significant results was found for the Spearman correlation and binomial sign tests. When we repeated the analysis for the other diseases under study, we did not observe significant correlations or binomial sign test results. The significance of this observation is unknown. Wagner et al.[33] reported complex relationships between methylation and expression, and that the strongest correlation was found in a group of developmentally important genes. Interestingly, schizophrenia is well-known to be a neurodevelopmental disorder and the stronger methylation-expression link is broadly consistent with Wagner et al. It is worth noting that the numbers of top methylation loci were small (<10) for CAD, DM and AD, and these samples might be underpowered to detect any correlations.

**Discussion**

In summary, we propose a general framework for epigenome-wide association study and integrative analysis with the transcriptome using GWAS summary statistics. The framework is based on the Mendelian randomization methodology, adapted to a methylome-wide scale. By applying the framework to five complex diseases, we showed how this approach may help to identify differentially methylated loci and prioritize genes for further functional or other follow-up studies. In addition, we observed strong evidence for differential expression among the top genes mapped to methylation loci.

There are some limitations to this study. Firstly, as alluded to earlier, the current method does not prove causal relationships. It is difficult to differentiate pleiotropy from causation, especially with only a few number of instrumental genetic variants. Recently methods have been proposed to correct the bias due to pleiotropy[13], but in the presence of few instrumental variables the results may not be reliable. Large-scale meQTL studies may enable more variants to be discovered and such bias-correction methods can potentially be applied. However, even if the differential methylation is due to pleiotropy, the associated gene or genetic variant might still represent a biologically important candidate for further studies. Conventional EWAS are not immune to associations due to pleiotropic effects either. Also, as the MR approach is less affected by confounding or reverse causation, the results from the MR-based EWAS can be used to prioritize findings from conventional EWAS for finding casual loci.

Another limitation is the tissue specificity. We used methylation and expression in the peripheral blood as a surrogate in this study. Obviously, the framework presented here is readily applicable to any tissue types, if relevant summary statistics are available. Summary data for meQTLs are less readily available compared to eQTLs. Large-scale resources like GTEx[36] which provides public access to eQTL information over a large range of tissues are still lacking for methylation. In addition, as argued by Zhu et al.[12], the exact tissue type implicated in a disease may be unknown. For example, psychiatric disorders obviously involve the brain but there is growing evidence for the immune system to be implicated[37,38].

The ability to detect associations with the MR approach will depend on the power of the original GWAS and QTL study. The method will become more powerful with increases in sample sizes. It is worth mentioning that the ARIES dataset only included females when studying methylation levels at middle age, and how well it can be generalized to adult males remain to be investigated.

Notwithstanding the above limitations, we believe the framework proposed here will be useful for researchers to gain insight into disease mechanisms, which might ultimately lead to improved treatment for complex diseases. This work also illustrates the potential of employing publicly available data, including large-scale GWAS and QTL studies, in integrating omics sources to achieve more in-depth understanding of complex traits and diseases.

**Acknowledgements**

This work is partially supported by the Lo-Kwee Seong Biomedical Research Fund and a CUHK direct grant. We thank Mr. Carlos Chau for data handling. We thank the Hong Kong Bioinformatics Center at CUHK for computing support. The author declares no competing interests.

Table 1 Methylation probes with mapped genes passing Bonferroni correction and *not* achieving genome-wide significance in the original GWAS meta-analysis

| Methylation Probe (time-point) | cpg_chr | cpg_pos | cpg_gene | nsnp | b.fixed.delta | pval.fixed.delta |
|---|---|---|---|---|---|---|
| **SCZ** | | | | | | |
| cg26536240 (Adolescence) | 16 | 89509760 | *ANKRD11* | 1 | 0.092 | 1.12E-06 |
| cg16263627 (Childhood) | 15 | 100339283 | *DNM1P46* | 1 | -0.206 | 1.52E-07 |
| cg21290290 (Childhood) | 1 | 2518275 | *FAM213B* | 1 | 0.233 | 1.95E-06 |
| cg08619954 (Childhood) | 4 | 5526783 | *LINC01587* | 1 | 0.157 | 9.89E-07 |
| cg16492833 (Adolescence) | 6 | 33679775 | *UQCC2* | 2 | 0.092 | 1.19E-06 |
| cg00658411 (Childhood) | 16 | 4467558 | *CORO7* | 1 | 0.063 | 1.16E-06 |
| cg14092988 (Adolescence) | 3 | 52407081 | *DNAH1* | 1 | -0.065 | 3.66E-07 |
| cg06606381 (Adolescence) | 12 | 133084897 | *FBRSL1* | 1 | 0.429 | 1.04E-13 |
| cg23656755 (Adolescence) | 1 | 230203043 | *GALNT2* | 1 | -0.201 | 2.38E-07 |
| cg10639394 (Childhood) | 15 | 82640457 | *GOLGA6L10* | 2 | 0.171 | 1.09E-07 |
| cg01061553 (Adolescence) | 12 | 110891467 | *GPN3* | 1 | -0.120 | 1.85E-06 |
| cg18278486 (Adolescence) | 6 | 26987575 | *LOC100270746* | 2 | 0.127 | 4.53E-08 |
| cg03959625 (Childhood) | 15 | 84868606 | *GOLGA2P7* | 1 | -0.167 | 5.39E-08 |
| cg19630374 (Adolescence) | 17 | 18023558 | *MYO15A* | 1 | -0.126 | 9.82E-07 |
| cg05725404 (Childhood) | 16 | 58534157 | *NDRG4* | 3 | -0.118 | 3.17E-09 |
| cg17372223 (Childhood) | 3 | 52568218 | *NT5DC2* | 2 | -0.089 | 1.15E-07 |
| cg02753903 (Childhood) | 19 | 38943525 | *RYR1* | 1 | -0.101 | 7.65E-07 |
| cg24247370 (Birth) | 13 | 99142703 | *STK24* | 1 | 0.120 | 4.00E-07 |
| cg18393722 (Birth) | 15 | 85113863 | *UBE2QP1* | 3 | -0.100 | 4.46E-07 |
| **AD** | | | | | | |
| cg18816397 (Middle age) | 6 | 32489555 | *HLA-DRB5* | 2 | 0.154 | 7.51E-10 |
| cg15224348 (Middle age) | 19 | 45543538 | *SFRS16* | 1 | 0.114 | 9.19E-08 |
| cg10853231 (Adolescence) | 19 | 48111307 | *GLTSCR1* | 1 | -0.131 | 2.05E-06 |
| **CAD** | | | | | | |
| cg25035485 (Childhood) | 2 | 21266500 | *APOB* | 2 | 0.132 | 8.45E-09 |
| cg03493300 (Childhood) | 10 | 104813866 | *CNNM2* | 1 | -0.063 | 2.17E-07 |
| cg25313468 (Adolescence) | 4 | 57773308 | *REST* | 2 | -0.106 | 2.32E-07 |
| cg24147428 (Childhood) | 11 | 65409760 | *SIPA1* | 2 | 0.098 | 2.49E-07 |
| cg05180856 (Birth) | 15 | 91428056 | *FES* | 1 | 0.167 | 9.29E-07 |
| **DM** | | | | | | |
| cg11823178 (Adolescence) | 8 | 41519399 | *ANK1* | 1 | 0.118 | 7.83E-07 |
| cg00908766 (Middle age) | 1 | 109817496 | *CELSR2* | 2 | -0.093 | 2.31E-06 |
| cg11879188 (Childhood) | 9 | 136149908 | *ABO* | 2 | 0.131 | 4.30E-06 |
| **IBD** | | | | | | |
| cg18477569 (Birth) | 6 | 90951134 | *BACH2* | 1 | -0.096 | 1.32E-06 |
| cg09501687 (Childhood) | 5 | 150169781 | *SMIM3* | 1 | 0.239 | 8.28E-09 |
| cg21582582 (Birth) | 3 | 182698605 | *DCUN1D1* | 1 | -0.186 | 3.89E-10 |
| cg10154826 (Birth) | 6 | 17600994 | *FAM8A1* | 1 | -0.209 | 1.13E-24 |
| cg14895029 (Childhood) | 7 | 2775587 | *GNA12* | 2 | 0.115 | 1.40E-06 |
| cg23982607 (Adolescence) | 1 | 1823379 | *GNB1* | 3 | 0.107 | 2.96E-07 |
| cg23216724 (Childhood) | 6 | 167571584 | *GPR31* | 1 | 0.247 | 6.88E-07 |
| cg23283495 (Adolescence) | 1 | 209979779 | *IRF6* | 3 | 0.120 | 6.00E-09 |
| cg01311341 (Adolescence) | 22 | 25575246 | *KIAA1671* | 1 | 0.255 | 1.03E-07 |
| cg21574349 (Adolescence) | 3 | 185080951 | *MAP3K13* | 1 | -0.308 | 2.14E-08 |
| cg00580354 (Adolescence) | 2 | 179296317 | *PRKRA* | 1 | -0.387 | 1.95E-19 |
| cg04092800 (Childhood) | 5 | 40681444 | *PTGER4* | 1 | 0.469 | 7.00E-09 |
| cg00271210 (Childhood) | 6 | 167070053 | *RPS6KA2* | 1 | -0.132 | 2.89E-10 |
| cg01358966 (Adolescence) | 16 | 29118931 | *RRN3P2* | 1 | 0.348 | 1.18E-06 |
| cg13746518 (Adolescence) | 9 | 137332241 | *RXRA* | 1 | -0.350 | 5.98E-11 |
| cg15335139 (Childhood) | 3 | 50242325 | *SLC38A3* | 1 | -0.114 | 8.34E-07 |
| cg22887911 (Childhood) | 3 | 44903730 | *TMEM42* | 1 | -0.262 | 2.52E-09 |
| cg10500283 (Adolescence) | 1 | 7913081 | *UTS2* | 2 | 0.131 | 1.98E-07 |
| cg21665057 (Adolescence) | 3 | 196295764 | *WDR53* | 1 | -0.195 | 6.67E-10 |

Only methylation probes with mapped genes are shown. We used Biomart to map genome-wide significant SNPs to genes; in this table only the "new" associations that do not have a GWAS hit are shown. cpg_chr, chromosome of the CpG probe, cpg_pos, position of the CpG probe; nsnp, number of SNPs acting as instrumental variables; b.fixed.delta, regression coefficient (effect size) from MR analysis using a fixed-effects model with variance estimated by the delta method; pval.fixed.delta, corresponding p-value from MR analysis.

**Table 2** All genes passing Bonferroni correction for the conjunction test of both differential methylation and expression (using maximum p)

| Cpg_gene | Simes_methyl_p | Expr_p | Total_cpg | Max_conj_p | Cpg_gene | Simes_methyl_p | Expr_p | Total_cpg | Max_conj_p |
|---|---|---|---|---|---|---|---|---|---|
| **SCZ** | | | | | **IBD** | | | | |
| *BTN3A2* | 7.89E-10 | 1.43E-18 | 1 | 7.89E-10 | *HLA-DQB2* | 6.43E-13 | 2.98E-15 | 3 | 6.43E-13 |
| *NMB* | 2.49E-09 | 1.94E-09 | 4 | 2.49E-09 | *ORMDL3* | 1.23E-12 | 1.48E-12 | 4 | 1.48E-12 |
| *ARL6IP4* | 7.19E-10 | 5.15E-09 | 3 | 5.15E-09 | *ZGPAT* | 4.28E-11 | 1.35E-14 | 3 | 4.28E-11 |
| *NDRG4* | 1.90E-08 | 3.08E-09 | 6 | 1.90E-08 | *SLC22A5* | 8.96E-11 | 1.52E-12 | 2 | 8.96E-11 |
| *KLC1* | 5.89E-08 | 8.83E-12 | 2 | 5.89E-08 | *INPP5E* | 1.38E-10 | 2.25E-10 | 1 | 2.25E-10 |
| *PSMA4* | 3.55E-08 | 6.15E-08 | 1 | 6.15E-08 | *APEH* | 3.08E-10 | 4.02E-20 | 7 | 3.08E-10 |
| *RPRD2* | 1.14E-09 | 1.19E-07 | 1 | 1.19E-07 | *STMN3* | 7.35E-10 | 1.94E-12 | 5 | 7.35E-10 |
| *ABCB9* | 1.25E-07 | 1.23E-10 | 1 | 1.25E-07 | *LIME1* | 4.43E-09 | 8.13E-16 | 6 | 4.43E-09 |
| *RERE* | 8.45E-08 | 3.51E-07 | 5 | 3.51E-07 | *PTGER4* | 7.04E-09 | 1.32E-10 | 1 | 7.04E-09 |
| *ITIH4* | 7.68E-07 | 6.79E-07 | 1 | 7.68E-07 | *RNASET2* | 5.22E-09 | 7.60E-09 | 2 | 7.60E-09 |
| *SNX19* | 1.38E-06 | 5.60E-11 | 2 | 1.38E-06 | *IKZF3* | 1.08E-09 | 1.43E-08 | 1 | 1.43E-08 |
| *GATAD2A* | 1.52E-06 | 1.42E-07 | 3 | 1.52E-06 | *ARPC2* | 1.55E-07 | 3.84E-08 | 2 | 1.55E-07 |
| *GPN3* | 1.85E-06 | 1.37E-06 | 1 | 1.85E-06 | *GSDMB* | 1.87E-07 | 8.27E-14 | 1 | 1.87E-07 |
| *CYP2D6* | 1.92E-06 | 1.01E-07 | 1 | 1.92E-06 | *ZGLP1* | 2.28E-07 | 2.27E-13 | 1 | 2.28E-07 |
| *FES* | 6.13E-07 | 2.01E-06 | 1 | 2.01E-06 | *UBE2L3* | 2.42E-07 | 2.35E-09 | 1 | 2.42E-07 |
| *NEK4* | 2.44E-06 | 7.35E-11 | 1 | 2.44E-06 | *PGAP3* | 7.00E-07 | 5.34E-08 | 2 | 7.00E-07 |
| *PITPNM2* | 4.74E-06 | 6.64E-06 | 2 | 6.64E-06 | *CUL2* | 1.67E-06 | 4.86E-11 | 1 | 1.67E-06 |
| *ZSCAN16* | 6.82E-06 | 4.64E-22 | 2 | 6.82E-06 | *IL1RL1* | 2.45E-06 | 1.03E-09 | 1 | 2.45E-06 |
| *NAGA* | 7.41E-06 | 3.31E-08 | 1 | 7.41E-06 | *IL27* | 9.05E-08 | 3.20E-06 | 1 | 3.20E-06 |
| *C22orf32* | 7.59E-06 | 3.60E-07 | 2 | 7.59E-06 | *GNB1* | 5.92E-07 | 3.58E-06 | 2 | 3.58E-06 |
| *SH3RF1* | 8.29E-06 | 1.19E-07 | 2 | 8.29E-06 | *CXCR2* | 3.77E-06 | 5.70E-07 | 1 | 3.77E-06 |
| | | | | | *SMURF1* | 4.09E-06 | 3.26E-07 | 1 | 4.09E-06 |
| **AD** | | | | | *TNFSF8* | 3.64E-06 | 4.53E-06 | 1 | 4.53E-06 |
| *CD2AP* | 2.00E-07 | 3.55E-09 | 1 | 2.00E-07 | *SULT1A2* | 5.34E-06 | 2.00E-10 | 2 | 5.34E-06 |
| *BIN1* | 9.10E-07 | 1.41E-07 | 2 | 9.10E-07 | *RBM6* | 5.38E-06 | 3.96E-07 | 1 | 5.38E-06 |
| | | | | | *IRF6* | 2.47E-06 | 6.15E-06 | 9 | 6.15E-06 |
| **CAD** | | | | | *STARD3* | 7.43E-06 | 2.53E-06 | 4 | 7.43E-06 |
| *LIPA* | 4.05E-12 | 3.74E-12 | 6 | 4.05E-12 | *RPS6KA2* | 6.36E-09 | 8.15E-06 | 22 | 8.15E-06 |
| *GGCX* | 5.18E-09 | 6.38E-09 | 1 | 6.38E-09 | *CARD9* | 8.92E-06 | 2.52E-25 | 2 | 8.92E-06 |
| *FES* | 9.29E-07 | 1.48E-07 | 1 | 9.29E-07 | | | | | |
| **DM** | | | | | | | | | |
| *TP53INP1* | 6.28E-09 | 1.10E-07 | 6 | 1.10E-07 | | | | | |

Cpg_gene: genes mapped to methylation loci; Simes_methyl_p, gene-based p-value from methylation using Simes test; Expr_p, expression p-values from MetaXcan; Total_cpg, total number of methylation lobes combined for this gene; Max_conj_p, conjunction test of differential expression and methylation by taking the maximum p.

Table 3 Top ten genes for each disease as ranked by their co-fdr

| Cpg_gene | Simes_methyl_p | Expr_p | Total_cpg | Simes_conj_p | Cofdr |
|---|---|---|---|---|---|
| **SCZ** | | | | | |
| ZSCAN16 | 6.96E-06 | 4.64E-22 | 2 | 9.28E-22 | 7.55E-16 |
| PRRT1 | 9.66E-03 | 1.72E-22 | 1 | 3.44E-22 | 1.09E-13 |
| BTN3A2 | 7.89E-10 | 1.43E-18 | 1 | 2.87E-18 | 1.37E-13 |
| NMB | 2.49E-09 | 1.94E-09 | 4 | 2.49E-09 | 1.71E-09 |
| KLC1 | 5.89E-08 | 8.83E-12 | 2 | 1.77E-11 | 1.60E-08 |
| BTN1A1 | 8.47E-01 | 5.18E-17 | 1 | 1.04E-16 | 2.04E-08 |
| ARL6IP4 | 3.75E-10 | 5.15E-09 | 3 | 7.49E-10 | 6.14E-08 |
| HIST1H3B | 5.83E-03 | 3.62E-15 | 1 | 7.23E-15 | 9.49E-08 |
| ABCB9 | 2.18E-07 | 1.23E-10 | 2 | 2.46E-10 | 1.45E-07 |
| SNX19 | 1.38E-06 | 5.60E-11 | 2 | 1.12E-10 | 2.68E-07 |
| **AD** | | | | | |
| TOMM40 | 8.56E-01 | 1.69E-29 | 1 | 3.38E-29 | 8.04E-24 |
| DMWD | 1.01E-03 | 2.76E-16 | 2 | 5.51E-16 | 1.05E-14 |
| CD2AP | 3.01E-07 | 3.55E-09 | 2 | 7.10E-09 | 8.01E-10 |
| BIN1 | 9.10E-07 | 1.41E-07 | 2 | 2.82E-07 | 9.87E-10 |
| MS4A6A | 4.88E-05 | 2.98E-10 | 1 | 5.97E-10 | 1.06E-08 |
| MYPOP | 5.74E-02 | 2.68E-11 | 2 | 5.35E-11 | 2.26E-08 |
| PNOC | 1.64E-02 | 1.23E-10 | 6 | 2.46E-10 | 3.25E-07 |
| DMPK | 1.21E-05 | 4.76E-06 | 2 | 9.52E-06 | 3.53E-07 |
| CLPTM1 | 3.95E-08 | 2.38E-03 | 1 | 7.90E-08 | 3.58E-07 |
| MS4A4A | 2.18E-01 | 2.06E-10 | 1 | 4.12E-10 | 5.57E-07 |
| **CAD** | | | | | |
| LIPA | 4.05E-12 | 3.74E-12 | 6 | 4.05E-12 | 1.07E-19 |
| CELSR2 | 9.38E-17 | 3.25E-04 | 4 | 1.88E-16 | 1.86E-16 |
| PSRC1 | 1.17E-04 | 8.21E-19 | 1 | 1.64E-18 | 2.20E-15 |
| GGCX | 5.18E-09 | 6.38E-09 | 1 | 6.38E-09 | 2.15E-13 |
| FES | 1.05E-06 | 1.48E-07 | 1 | 2.96E-07 | 3.81E-08 |
| SIPA1 | 9.97E-07 | 2.35E-05 | 4 | 1.99E-06 | 1.55E-07 |
| SLC22A3 | 4.20E-04 | 1.15E-07 | 1 | 2.30E-07 | 5.46E-07 |
| DHX36 | 6.01E-05 | 2.10E-05 | 2 | 4.20E-05 | 8.60E-06 |
| TAGLN2 | 4.37E-06 | 3.38E-05 | 1 | 8.73E-06 | 1.08E-05 |
| SH2B3 | 2.57E-05 | 1.41E-05 | 1 | 2.57E-05 | 2.04E-05 |

| Gene | Simes_methyl_p | Expr_p | Total_cpg | Simes_conj_p | Cofdr |
|---|---|---|---|---|---|
| **DM** | | | | | |
| *TP53INP1* | 6.28E-09 | 1.10E-07 | 6 | 1.26E-08 | 3.58E-10 |
| *JAZF1* | 5.52E-08 | 3.49E-03 | 1 | 1.10E-07 | 1.84E-04 |
| *CAMK1D* | 6.61E-01 | 1.61E-06 | 1 | 3.21E-06 | 3.48E-04 |
| *PCBD2* | 2.18E-05 | 4.93E-04 | 1 | 4.36E-05 | 4.24E-04 |
| *UTS2* | 3.05E-02 | 9.61E-06 | 2 | 1.92E-05 | 5.55E-04 |
| *ABCB9* | 9.52E-05 | 8.68E-05 | 1 | 9.52E-05 | 1.09E-03 |
| *UBE2E2* | 1.86E-06 | 1.64E-03 | 1 | 3.72E-06 | 2.58E-03 |
| *KCNK17* | 1.93E-04 | 2.12E-04 | 2 | 2.12E-04 | 2.91E-03 |
| *ANK1* | 4.35E-06 | 5.54E-03 | 4 | 8.71E-06 | 3.00E-03 |
| *COPA* | 1.10E-03 | 9.49E-04 | 1 | 1.10E-03 | 2.00E-02 |
| **IBD** | | | | | |
| *HLA-DRB5* | 1.15E-31 | 1.03E-01 | 4 | 2.29E-31 | 2.11E-24 |
| *CARD9* | 8.92E-06 | 2.52E-25 | 2 | 5.05E-25 | 1.32E-20 |
| *HLA-DQB2* | 6.43E-13 | 2.98E-15 | 3 | 5.96E-15 | 5.37E-19 |
| *FAM8A1* | 1.13E-24 | 3.08E-01 | 1 | 2.27E-24 | 1.93E-18 |
| *APEH* | 3.08E-10 | 4.02E-20 | 7 | 8.04E-20 | 2.65E-18 |
| *LIME1* | 4.43E-09 | 8.13E-16 | 6 | 1.63E-15 | 2.06E-16 |
| *ZGPAT* | 4.28E-11 | 1.35E-14 | 3 | 2.71E-14 | 3.58E-15 |
| *ORMDL3* | 1.23E-12 | 1.48E-12 | 4 | 1.48E-12 | 5.41E-15 |
| *SLC22A5* | 8.96E-11 | 1.52E-12 | 2 | 3.04E-12 | 3.14E-13 |
| *STMN3* | 9.98E-09 | 1.94E-12 | 5 | 3.89E-12 | 3.50E-13 |

The null hypothesis is the global null, i.e. the gene is not associated with methylation or expression changes. Cpg_gene: genes mapped to methylation loci; Simes_methyl_p, gene-based p-value from methylation using Simes test; Expr_p, expression p-values from MetaXcan; Total_cpg, total number of methylation lobes combined for this gene; Simes_conj_p, p-value by Simes test (testing the global null that the gene is not associated with methylation or expression changes); Cofdr, co-false discovery rate with respect to the global null.

Table 4  Differential expression within top methylation loci

|  | Fisher (diff expr) | Simes (diff expr) |
|---|---|---|
| SCZ | 5.95E-113 | 5.30E-17 |
| AD | 1.20E-14 | 1.20E-14 |
| CAD | 4.68E-27 | 3.36E-11 |
| DM | 2.09E-11 | 5.48E-07 |
| IBD | 2.01E-184 | 2.09E-18 |

Fisher (diff expr), Fisher's method for combining expression p-values from top methylation loci; Simes (diff expr), Simes method for combining expression p-values.

Table 5  Correlation between effect sizes of differential methylation and expression

|  |  | SCZ | AD | CAD | DM | IBD |
|---|---|---|---|---|---|---|
| Overall | Pearson cor | -0.571 | 0.462 | 0.065 | 0.200 | -0.065 |
|  | Pearson cor p | 1.79e-4 | 0.211 | 0.850 | 0.704 | 0.619 |
|  | Spearman cor | -0.522 | 0.527 | -0.018 | -0.290 | 0.093 |
|  | Spearman cor p | 7.74e-4 | 0.145 | 0.957 | 0.577 | 0.475 |
|  | Opp sign/total | 27/38 | 2/9 | 4/11 | 2/6 | 28/61 |
|  | Binomial test p | 0.014 | 0.180 | 0.549 | 0.688 | 0.609 |
| Promotor region | Promoter_region cor | -0.611 | 0.310 | -0.352 | 0.200 | -0.017 |
|  | Promoter_region cor p | 5.46e-3 | 0.690 | 0.648 | 0.800 | 0.926 |
|  | Spearman cor | -0.630 | -0.200 | -0.632 | -0.400 | 0.106 |
|  | Spearman cor p | 4.71e-3 | 0.917 | 0.368 | 0.750 | 0.556 |
|  | Opp sign/total | 15/19 | 1/4 | 2/4 | 2/4 | 16/33 |
|  | Binomial test p (one-tailed) | 0.010 | 0.938 | 0.688 | 0.688 | 0.636 |
|  | Binomial test p (two-tailed) | 0.019 | 0.625 | 1 | 1 | 1 |
| Non-promoter region | Other regions cor | -0.540 | 0.650 | 0.475 | NA | -0.082 |
|  | Other regions cor p | 0.017 | 0.235 | 0.281 | NA | 0.677 |
|  | Spearman cor | -0.246 | 0.900 | 0.595 | NA | 0.016 |
|  | Spearman cor p | 0.309 | 0.083 | 0.250 | NA | 0.936 |
|  | Opp sign/total | 12/19 | 1/5 | 2/7 | 0/2 | 12/28 |
|  | Binomial test p | 0.359 | 0.375 | 0.453 | 0.500 | 0.572 |

Methylation probes are grouped by genes and their locations for computing correlations. Effect sizes are averaged if a gene is represented by multiple methylation probes. Overall cor, overall correlation regardless of locations; cor, correlation; Opp sign/total, effect sizes with opposite signs/total; Binomial test *p* (one-tailed), one-tailed binomial test for opposite direction of effect sizes.